\documentclass{sig-alternate}
\usepackage{times}
\usepackage{comment}
\usepackage{soul}
\usepackage[T1]{fontenc}
\usepackage[labelfont=bf]{caption} % Make figure numbering in captions bold

\usepackage{MnSymbol}
\usepackage{multirow}
\usepackage{pgfplots}
\usepackage{subfigure}
\usepackage{graphicx,color}
\usepackage{colortbl}
\usepackage{caption}
\DeclareCaptionType{copyrightbox}
\usepackage[utf8]{inputenc}
\usepackage{xspace}
\usepackage{url}
\usepackage{semantic}
\usepackage[nocompress]{cite}
\usepackage{calc}
\usepackage{paralist}

\usepackage{booktabs}
\usepackage{multirow}

\title{Towards Lifecycle Management of Collaborative Analysis Workflows through Provenance Capture and Analysis}
\title{Lifecycle Management of Collaborative Analysis Workflows}
\title{ProvDB: A System for Lifecycle Management of Collaborative Analysis Workflows}

\author{
Hui Miao,~Amit Chavan,~ Amol Deshpande\\
       \affaddr{University of Maryland, College Park, MD, USA}\\
       \email{{\large{\{}}hui,~amitc,~amol{\large{\}}}@cs.umd.edu} 
}

\newcommand{\topic}[1]{\par \smallskip \noindent{\bf {#1}}}
\newcommand{\topicul}[1]{\par \smallskip \noindent\underline{\bf {#1}}}

\newcommand{\gdoccomment}[1]{\texttt{\small{\color{gray}{ Gdoc: #1}}}}
\renewcommand{\gdoccomment}[1]{}
\newcommand{\huicomment}[1]{\texttt{\small{\color{blue}{ Hui: #1}}}}
\renewcommand{\huicomment}[1]{}

\newcounter{cctr}
\stepcounter{cctr}

\newcounter{tctr}
\stepcounter{tctr}

\newcommand{\squishlist}{
   \begin{list}{$\bullet$}
    { \setlength{\itemsep}{0pt}
      \setlength{\parsep}{2pt}
      \setlength{\topsep}{6pt}
      \setlength{\partopsep}{0pt}
      \leftmargin=25pt
\rightmargin=0pt
\labelsep=5pt
\labelwidth=10pt
\itemindent=0pt
\listparindent=0pt
\itemsep=\parsep
    }
}
\newcommand{\squishend}{\end{list}}

\renewenvironment{description}[1][0pt]
  {\renewcommand\descriptionlabel[1]{\underline{\emph{##1:}} }
  \list{}{\labelwidth=0pt \leftmargin=#1 \itemsep=2pt \parsep=0pt \topsep=3pt \partopsep=0pt \rightmargin=0pt \labelsep=0pt \itemindent=0pt \listparindent=\parindent
   }}
  {\endlist}

\newcommand{\eat}[1]{}

\newcommand{\nosub}[1]{#1}

\newcommand{\provdb}{{\sc ProvDB}\xspace}
\newcommand{\git}{{\tt git}\xspace}

\begin{document}
\maketitle

\begin{abstract}
As data-driven methods are becoming pervasive in a wide variety of disciplines, there is an urgent need to develop scalable and sustainable tools to simplify the process of data science, to make it easier to keep track of the analyses being performed and datasets being generated, and to enable introspection of the workflows. In this paper, we describe our vision of a unified provenance and metadata management system to support lifecycle management of complex collaborative data science workflows. We argue that a large amount of information about the analysis processes and data artifacts can, and should be, captured in a semi-passive manner; and we show that querying and analyzing this information can not only simplify bookkeeping and debugging tasks for data analysts but can also enable a rich new set of capabilities like identifying flaws in the data science process itself. It can also significantly reduce the time spent in fixing post-deployment problems through automated analysis and monitoring. We have implemented an initial prototype of our system, called \provdb, on top of {\tt git} (a version control system) and Neo4j (a graph database), and we describe its key features and capabilities.
\end{abstract}

\section{Introduction}

Data-driven methods and products are becoming increasingly common in a variety of communities, including sciences, education, economics, and social and web analytics. This has resulted in a pressing need for sustainable and scalable tools that facilitate the {\em end-to-end data science process} by making it easy to maintain and share time-evolving datasets; to collaboratively clean, integrate, and analyze datasets; to perform introspective analysis to identify errors in the data science pipelines; and to learn from other data scientists. This is especially challenging because the process of collaborative data science is often ad hoc, typically featuring highly unstructured datasets, an amalgamation of different tools and techniques, significant back-and-forth among the members of a team, and trial-and-error to identify the right analysis tools, algorithms, models, and parameters.  Although there is much prior and ongoing work on developing tools to perform specific data analysis tasks, platform support for managing the end-to-end process is still lacking in practice. There is no easy way to capture and reason about ad hoc data science pipelines, many of which are often spread across a collection of analysis scripts. Metadata or provenance information about how datasets were generated, including the programs or scripts used for generating them and/or values of any crucial parameters, is often lost. Similarly, it is hard to keep track of any dependencies between the datasets. As most datasets and analysis scripts evolve over time, there is also a need to keep track of their {\em versions} over time; using version control systems (VCS) like {\em git} can help to some extent, but those don't provide sufficiently rich introspection capabilities.

Lacking platform support for capturing and analyzing such provenance and metadata information, data scientists are required to manually track and act upon such information, which is not only tedious, but error-prone. For example, data scientists must manually keep track of which derived datasets need to be updated when a source dataset changes. They often use spreadsheets to list which parameter combinations have been tried out during the development of  a machine learning model. Debugging becomes significantly harder; e.g., a small change in an analysis script may have significant impact on the final result, but identifying that change may be non-trivial, especially in a collaborative setting. It is similarly challenging to identify which input records are most relevant to a particular output record.  ``Repeatability'' can often be very difficult, even for the same researcher, because of an amalgamation of constantly evolving tools and datasets being used, and because of a lack of easy-to-use mechanism to keep track of the parameter values used during analysis or modeling. Critical errors may be hidden in the mess of datasets and analysis scripts that cannot be easily identified; e.g., a data scientist may erroneously be training on the test dataset due to an inadvertent mistake while creating the testing and training datasets.

This paper describes a system, called \provdb, for unified management of all kinds of metadata about collaborative data science workflows that gets generated during the analysis process; this includes (a) version lineages of data, scripts, and results (collectively called {\em artifacts}), (b) data provenance among artifacts which may or may not be structured, (c) workflow metadata on derivations and dependencies among artifact snapshots, and so on.  \ul{Our hypothesis is that by combining all this information in one place, and making it easy to analyze or query this information, we can enable a rich set of functionality that can simplify the lives of data scientists, make it easier to identify and eliminate errors, and decrease the time to obtain actionable insights.} This is hardly a new observation, and there has been much prior work on capturing and analyzing provenance in a variety of communities. However, there is still a lack of practical systems that treat different kinds of provenance and metadata information in a unified manner, and that can be easily integrated in the workflow of a data science project. \ul{At the same time, the widespread use of data science has brought to the forefront several important and crucial challenges, such as ethics, transparency, reproducibility, etc., and we posit that fine-granularity provenance is a key to addressing those challenges.} 
%In our work so far, we have identified some such potential opportunities as we discuss later in the paper.

\begin{comment}
We discuss the potential opportunities in exploiting such information and the crucial systems and conceptual challenges that need to be addressed, and describe our initial prototype implementation built 
on top of \git.
\end{comment}

%%% As above, should shorten as required
There are however several crucial systems and conceptual challenges that must be addressed to fully exploit those opportunities. Next, we briefly discuss those challenges, and how we address them in our prototype implementation, called \provdb. First, it is hard to define a {\em schema} for the provenance/metadata information a priori, and different users or different workflows may wish to capture and analyze different types of such data. Instead of requiring a specific schema, we advocate a ``schema-later'' approach, where a small base schema is fixed, but users can add arbitrary semistructured information (in JSON or equivalent formats) for recording additional metadata. The specific data model we use generalizes and refines a data model proposed in our prior work~\cite{provenance2015}, and allows flexibly capturing a variety of different types of information including versioning and provenance information, parameters used during experiments or modeling, statistics gathered to make decision, analysis scripts, notes or tags, etc. In our prototype implementation, we map this logical data model to a {\em property graph data model}, and use the Neo4j graph database to store the information.
\begin{comment}; it allows capturing the variety of different types of data and metadata commonly found in a broad range of data science applications, including versioning and provenance information, derivation information, parameters used during experiments or modeling, statistics gathered to make decisions, analysis scripts, notes or tags, etc. 
\end{comment}

Second, we must be able to capture the information with minimal involvement from the users, otherwise the system is unlikely to be used in practice. To address this, \provdb features a suite of extensible {\em provenance ingestion} mechanisms. The currently supported mechanisms target the scenario where the user primarily interacts with the system using a {\tt UNIX Shell}. The {\tt shell} commands run by the users to manipulate files or datasets are intercepted, and analyzed using a collection of {\em  ingestors}. These ingestors can analyze the before- and after-state of the artifacts to generate rich metadata; several such ingestors are already supported and new ingestors can be easily registered for specific commands. For example, such an ingestion program is used to analyze log files generated by Caffe (a deep learning framework) and generate metadata about accuracy and loss metrics (for learned models) in a fine-grained manner. 
\provdb also supports the notion of {\em file views}, where users can define file transformations (either as shell commands or simplified SQL); this functionality not only simplifies some transformation tasks (e.g., specifying training/testing splits or sub-sampling), but also allows us to capture fine-grained record-level dependencies.
\begin{comment}
 (e.g., although our implementation doesn't currently support it, prior developed techniques for analyzing provenance of SQL queries could now be used to generate record-level provenance across datasets).  
\end{comment}

Third, and perhaps conceptually the most difficult, challenge is to develop declarative abstractions to make it easy to exploit this data. There are many potential ways to use such data including: (a) explanation queries where we are looking for origins of a piece of data, (b) introspection queries that attempt to identify flaws with the data science process (e.g., {\em p-value hacking}), (c) continuous monitoring to identify issues during deployment of a data science pipeline (e.g., {\em concept drifts} where a learned model doesn't fit new data; changes to input data formats), and many others. We are working on developing and supporting a high-level DSL that enables a large range of such queries; however, formalizing some of these queries (e.g., identifying p-value hacking, or ethics issues) itself is a major challenge. Our prototype \provdb implementation features a web browser-based visualization tool for inspecting and querying the provenance information, that supports a collection of pre-defined queries; it also supports querying the information directly using Cypher, the Neo4j query language. It supports a limited form of {\em continuous monitoring}, where a user can specify a constraint to be monitored over a set of properties of artifacts.

Finally, we expect many efficiency and optimization issues that will arise as the volume of the captured data increases. This is especially expected to be an issue with the record-level provenance information; even the ``versioning'' information can be quite large because of what we call ``implicit'' versions, that are generated every time a provenance capture is initiated.

\provdb is being developed in conjunction with DataHub~\cite{datahub2015}, a dataset-centric platform for enabling collaborative data analytics that supports managing a large number of datasets, their versions over time, and derived data products. A prior paper~\cite{provenance2015} described an initial proposal for a query language for unified querying of provenance and versioning information, but did not have an implementation, and did not discuss the issues of how provenance information may be captured and the rich introspective analysis that may be performed on such information. The current \provdb prototype is built on top of \git, widely used by data scientists due to its intuitive support for collaboration, and Neo4j, a graph database system. As DataHub matures, we plan to integrate \provdb with it in future.

%\vspace{5pt}
\noindent{\bf Outline:} We begin with discussing closely related work and putting our work in context of that in Sec.~\ref{sec:related}. We then present \provdb system architecture (Sec.~\ref{sec:sys_arch}) and its data model (Sec.~\ref{sec:data_model}), followed by a discussion of provenance ingestion mechanisms that it supports (Sec.~\ref{sec:ingestion}). We then discuss the types of analyses that \provdb enables and the web browser-based visualization tool (Sec.~\ref{sec:query}).

\section{Prior Work}
\label{sec:related}

There has been much work on scientific workflow systems over the years, with some of the prominent systems being Kepler\nosub{~\cite{ludascher2006scientific}}, Taverna\nosub{~\cite{oinn2006taverna}}, Galaxy\nosub{~\cite{giardine2005galaxy}}, iPlant\nosub{~\cite{goff2011iplant}}, VisTrails\nosub{~\cite{bavoil2005vistrails}}, Chimera\nosub{~\cite{foster2002chimera}}, Pegasus\nosub{~\cite{DBLP:journals/sp/DeelmanSSBGKMVBGLJK05}}, to name a few
%\footnote{Please see: http://go.umd.edu/cidr for a detailed related work discussion with citations.}. 
These systems often center around creating, automating, and monitoring a well-defined workflow or data analysis pipeline. But they cannot easily handle fast-changing pipelines, and typically are not suitable for ad hoc collaborative data science workflows where clear established pipelines may not exist except in the final, stable versions. Moreover, these systems typically do not support the entire range of tools or systems that the users may want to use, they impose a high overhead on the user time and can substantially increase the development time, and often require using specific computational environment. Further, many of these systems require centralized storage and computation, which may not be an option for large datasets.

Many researchers find {\bf version control systems} (e.g., git, svn) and hosted platforms built around them (e.g., GitHub, GitLab) much more appropriate for their day-to-day needs. These systems provide transparent support for versioning and sharing, while imposing no constraints on what types of tools can be used for the data processing itself. Though these systems keep version lineage among committed artifacts, these systems are typically too ``low-level'', and have very little support for capturing higher-level workflows or for keeping track of the operations being performed or any kind of provenance information. The versioning API supported by these systems is based on a notion of files, and is not capable of allowing data researchers to reason about data contained within versions and the relationships between the versions in a holistic manner. 
Our proposed system can be seen as providing rich introspection and querying capabilities those systems lack. A wide range of analytic packages like SAS\nosub{~\cite{sas}}, Excel\nosub{~\cite{excel}}, R\nosub{~\cite{r}}, Matlab\nosub{~\cite{matlab}}, and Mahout\nosub{~\cite{mahout}}, or data science toolkits such as IPython\nosub{ \cite{ipython}}, Sci\-kit\nosub{~\cite{scikit}}, and Pandas\nosub{~\cite{pandas}}, are frequently used for performing analysis itself; however, those lack comprehensive data management or collaboration capabilities. 

There has been significant interest in developing general-purpose systems for handling different aspects of “model lifecycle management” in recent years. Much of that work has focused on the “training” aspect; several general-purpose systems like GraphLab, TensorFlow, Parameter Server, etc., have been designed over the years, and there is also much work on specific aspects of ML pipeline (e.g., feature engineering~\cite{columbus_tods16}. Several recent systems have attempted to address end-to-end issues, including model serving (e.g., TensorFlow Serving, Velox~\cite{verlox_cidr}, MSMS~\cite{model_selection_arun_record15}, ModelDB~\cite{modeldb_HILDA16}). Our work, to a large extent, is complementary to that work; our focus is on the lifecycle management when a data-driven project team consisting of data analysts/scientists at different skill levels are trying to collaboratively develop a model. \provdb can be used as the provenance data management layer for most such systems.

\huicomment{
We probably need to formally characterize lifecycle management system. More brainstorm is needed.
Random thought:
Lifecycle: ?understand->update->try out->compare->share->..; 
System Goal: accelerating lifecycle; 
Method: declarative construct, provenance, system enabling collaborative workflow;
Challenges: rich problems in those understand/update/try out/compare .. steps? ..
}

There has also been much work on {\bf provenance}, with increasing interest in the recent years. Provenance can be captured at different granularities, and at different levels of detail. In scientific workflow systems%~\cite{freire2008provenance}
\nosub{~\cite{freire2008provenance,simmhan2005survey,freire2008provenance,freire2006managing,altintas2006provenance,simmhan2006framework,ludascher2006scientific,oinn2006taverna,yu2005taxonomy,bavoil2005vistrails,foster2002chimera}} where the operations are typically treated as black boxes, the provenance can usually be captured only at the level of datasets. Workflow provenance may include: a) prospective information about the definition of the workflow, b) retrospective information during the execution of the workflow, c) metadata about blocks and datasets in a workflow, and d) input/output lineages among blocks~\cite{conf/ipaw/ZhaoWF06}.
On the other hand, in dataflow systems where the operators are written in a declarative language (e.g., SQL, Pig Latin, Spark), data provenance at record level can be captured if needed\nosub{~\cite{buneman2001and,trio,buneman2006provenance,green2007provenance,davidson2008provenance,DBLP:journals/ftdb/CheneyCT09,moreau2010foundations,sparkprovenance2015}}. 
Our work aims to combine the two together with version lineages captured by VCS, and provide uniform platform for collaborative data science workflows. Our work is complementary to, and can utilize, those prior techniques to capture the provenance information itself; our focus is primarily on how to exploit that information for providing richer introspection capabilities.

\nosub{
Several systems have been designed that focus on specific aspects of the proposed work, including collaborative data management  (e.g., Fusion tables \cite{DBLP:journals/debu/MadhavanBBGGHJLLLMSS12}, Orchestra~\cite{DBLP:conf/cidr/IvesKKC05}, CQMS~\cite{DBLP:conf/cidr/KhoussainovaBGKS09,DBLP:conf/ssdbm/HoweCSKKKB11}, LabBook~\cite{kandogan2015labbook}), data sharing (e.g., SQLShare~\cite{DBLP:conf/ssdbm/HoweCSKKKB11, DBLP:journals/debu/HoweH12,DBLP:conf/ssdbm/HalperinRWSHA13}, SMILE~\cite{smileedbt2014}), hosted data repositories accessible to applications (CKAN~\cite{ckan}, Domo~\cite{domo}), hosted data science (Domino~\cite{domino}), as well as data publishing tools  (Quandl~\cite{quandl}, Factual~\cite{factual}, DataMarket~\cite{data-market}). None of them, however, aim to capture a broad range of metadata and provenance information in a unified fashion, to support high-level analysis and reasoning over data science pipelines.}

LabBook~\cite{kandogan2015labbook} has somewhat similar goals, and also uses a property graph to manage metadata captured during collaborative analytics and features web-based apps architecture for analyzing the metadata. However, LabBook does not treat ``versioning'' as a first-class construct, and does not focus on developing passive provenance ingestion mechanisms or sophisticated querying abstractions as we do here. GOODS~\cite{halevy2016goods} is a dataset management system, that passively extracts metadata information from a large number of datasets within an enterprise, and allows users to find, monitor, and analyze datasets. Although it shares our philosophy of passive collection of metadata, the types of metadata collected and the analyses performed are vastly different.

\section{System Architecture}
\label{sec:sys_arch}

\gdoccomment{Merge some of the high-level discussion from the ModelHub paper. The basic architecture. How it is built on top of Git. Differences from that would be the standalone \provdb system. ``dlv'' now dumps the data in there, rather than storing it somewhere itself. The separate binary files stuff should also go away for now. }

\begin{figure}[t]             
%\vspace{-2pt}
\centering
\includegraphics[width=3.4in]{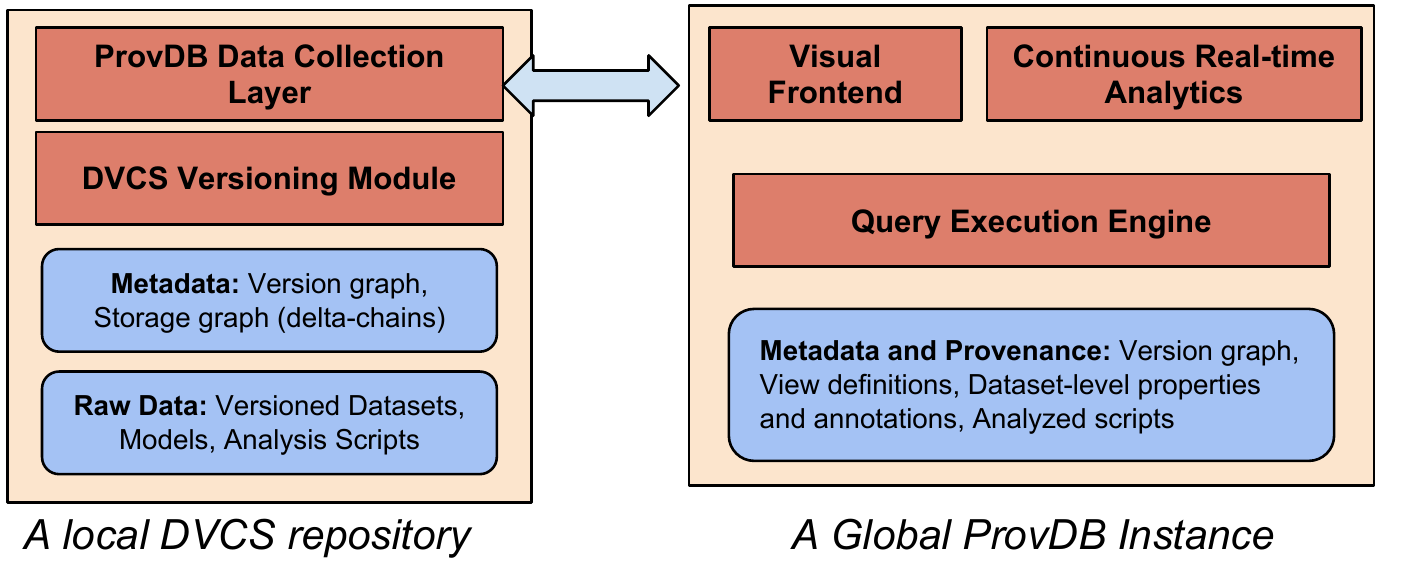}
%\vspace{-10pt}
 \caption{High-level System Architecture: a single \provdb instance will  manage the provenance and metadata information for a collaborative team, using a Neo4j instance to store data (not shown)}
%\vspace{-12pt}
\label{fig:arch}
\end{figure}

\provdb is a stand-alone system, designed to be used in conjunction with a dataset version control system (DVCS) like {\tt git} or {\tt DataHub} (Figure \ref{fig:arch}). The DVCS will handle the actual version management tasks, including supporting the standard {\em checkout}, {\em commit}, {\em merge}, etc., functionality, and the distributed and decentralized management of individual repositories.

We envision a number of local DVCS ``repositories'', each corresponding to a team of researchers collaborating closely together. The contents of each repository will typically be replicated across a number of machines as different researchers ``check out'' the repository contents to work on them. Since we leverage {\tt git} for keeping these in sync, the repository contents are available as files for the users to operate upon; the users can run whichever analysis tools they want on those after checking them out, including distributed toolkits like Hadoop or Spark.

A repository consists of a set of {\em versions}. A version, identified by an ID, is \ul{immutable} and any update to a version conceptually results in a new version with a different version ID (physical data structures are typically not immutable and the underlying DVCS uses various strategies for compact storage~\cite{dataset2015}). New versions can  also be created through the application of transformation programs to one or more existing versions. The version-level provenance that captures these processes is maintained as a ``version graph'', a directed acyclic graph with versions as nodes. Typically, the leaves of the version graph correspond to different {\em live branches} that different users may be operating upon at the same time. As we discuss in the next section, \provdb actually maintains a conceptual ``workflow graph'' with many other types of nodes and edges.
\begin{comment}
We assume the DVCS supports standard version control operations including {\sc checkout} a version or {\sc commit} a new version, create a {\sc branch}, {\sc merge} two or more versions, etc. 
On top of DVCS, \provdb intercepts user shell command, commits fine-grained versions, which maintained and forms a conceptual ``workflow graph'', where shell actions as nodes (derivations) and snapshots as edges (input/output). 
\end{comment}

Broadly, the data maintained across the system can be categorized into: (a) raw data that the users can directly access and analyze including the datasets, analysis scripts, and any derived artifacts such as trained models, and (b) metadata or provenance information transparently maintained by the system, and used for answering queries over the versioning or provenance information. Fine-grained record-level provenance information may or may not be directly accessible to the users depending on the ingest mechanism used. Note that, the split design that we have chosen to pursue requires duplication of some information in the DVCS and \provdb. We believe this is a small price to pay for the benefits of having a standalone provenance management system.

{\bf \provdb Data Collection Layer} is a thin layer on top of the DVCS (in our case, \git) that is used to capture the provenance and metadata information. This layer needs to support a variety of functionality to make it easy to collect a large amount of metadata and provenance information, with minimal overhead to the user (Sec.~\ref{sec:ingestion}). The \provdb instance itself is a separate process, and currently uses the Neo4j graph database to store the data; we chose Neo4j because of its support for the flexible property graph data model, and graph querying functionality out-of-the-box (Sec. \ref{sec:data_model}). The data stored inside \provdb can be accessed either through the Neo4j frontend, or through a visual frontend that we have built that supports a variety of provenance queries (Sec.~\ref{sec:query}).
\begin{comment}
All the data it collects will be stored in a distributed key-value store like Cassandra or a graph data management system like Titan (which itself uses Cassandra as a backend). Using a system like Titan gives us higher-level graph querying functionality out of the box, but at the same time, limits us to the supported {\em property graph} model. We plan to investigate both options in our work. 
\end{comment}

\begin{figure}[!t]
\centering
\includegraphics[width=0.45\textwidth]{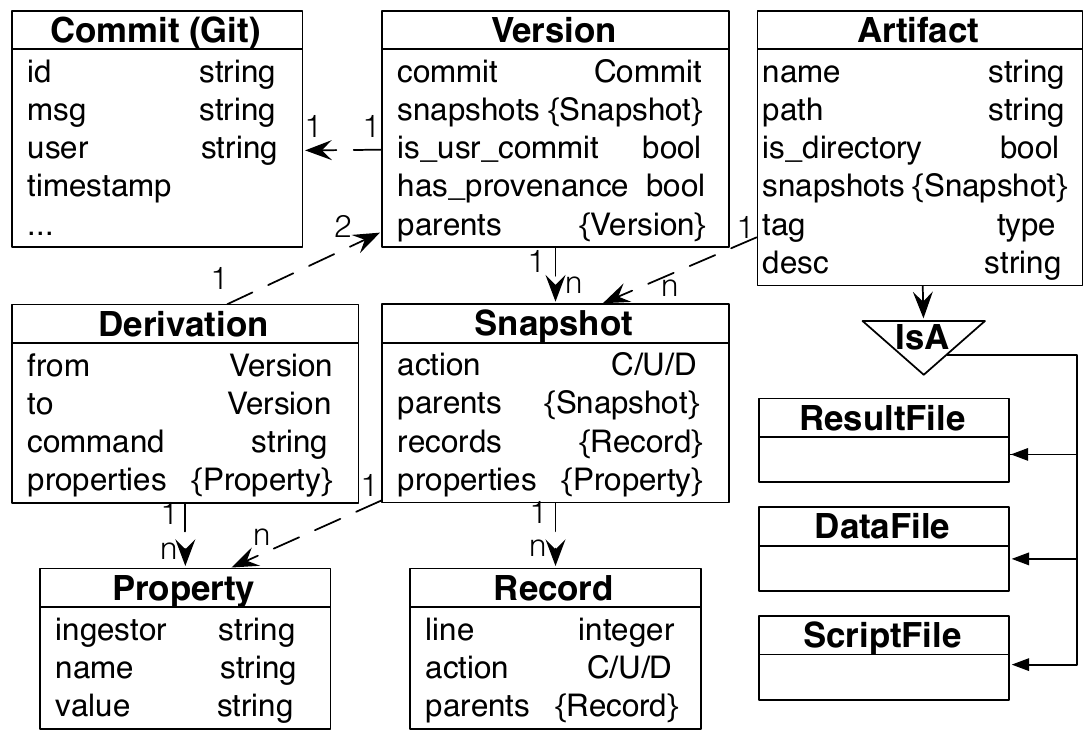}
%\vspace{-2pt}
\caption{Conceptual Data Model}
%\vspace{-15pt}
\label{fig:data_model}
\end{figure}

\begin{figure*}[t!]
%\subfigure[Conceptual Data Model]{
%  \includegraphics[width=0.5\textwidth]{fig_datamodel.pdf}
%  \label{fig:data_model}
%}
\centering{
\subfigure[Example Workflow]{
  \includegraphics[width=0.43\textwidth]{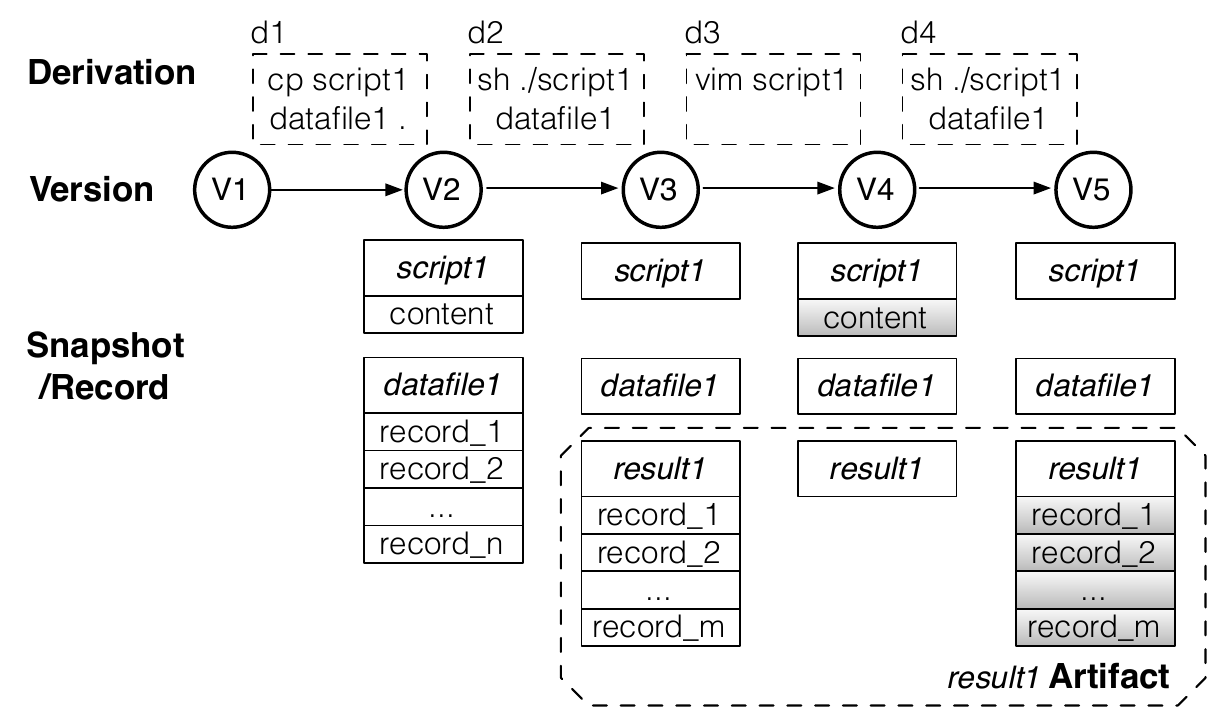}
  \label{fig:data_model_example}
}
~~
\subfigure[Provenance Property Graph]{
  \includegraphics[width=0.46\textwidth]{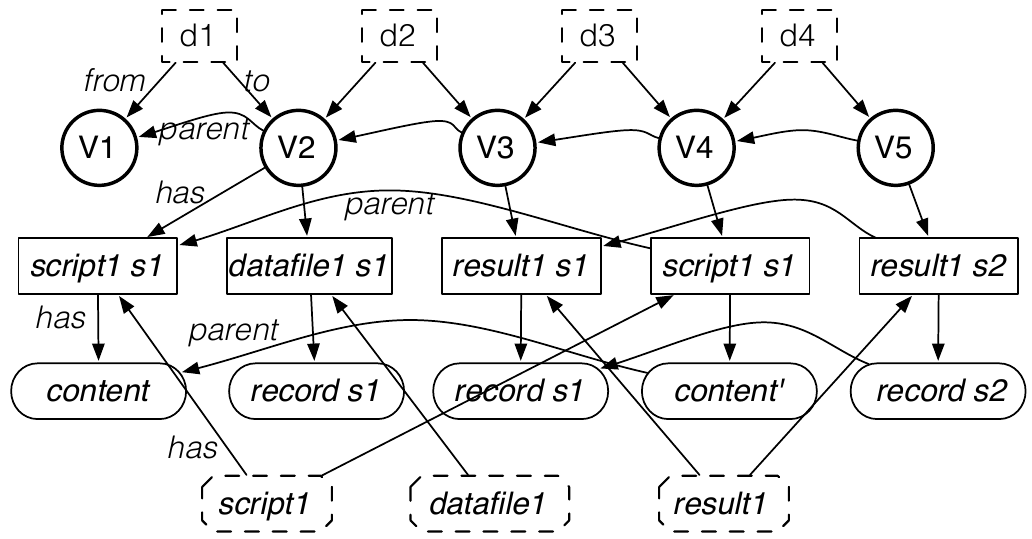}
  \label{fig:artifact_prov_graph}
}
}
%\vspace{-10pt}
\caption{An example workflow and the corresponding \provdb property graph}
%\vspace{-15pt}
\label{fig:data_model_all}
\end{figure*}

\section{Data model} 
\label{sec:data_model}

\gdoccomment{The data model will be a mix of what was described in the TAPP paper (below) and what is described in the ModelHub paper. However, described in the term of property graphs, assuming that's how we implement it.}

To encompass a large variety of situations, our goal was to have a flexible data model that reflects versioning and workflow pipelines, and supports addition of arbitrary metadata or provenance information. As such, we use fixed ``base schema'' (Figure \ref{fig:data_model}) to capture the information about the versions, the different data artifacts, and so on, while allowing arbitrary properties to be added to the various entities. We map this logical data model to a property graph model (Figure \ref{fig:artifact_prov_graph}), which we use as our physical data model. Our model differs from the other similar models proposed in the past work~\cite{conf/ipaw/ZhaoWF06,kandogan2015labbook} primarily in the explicit modeling of versions.

\begin{comment}In this section, we propose a data model to describe a collaborative data science project and enable provenance capturing and analysis at different levels. In order to have generality to be applied in various data science practice, we design the data model with minimal assumptions on the users. We do not assume the user aware of a DVCS (e.g. git) in the practice, also we do not assume her preferred development environment or task types. 
\end{comment}

\topicul{Conceptual Data Model}:
We view a data science project as a working directory with a set of {\bf artifacts} (files), and a development lifecycle as a series {\bf actions} (shell commands, edits, transformation programs) which perform create/read/update/delete (CRUD) operations in the working directory.

More specifically: an {\bf artifact} is a file, which the user modifies, runs, and talks about with peers. Artifacts can be tagged as belonging to one of three different types: \emph{ResultFile}, \emph{DataFile}, \emph{ScriptFile}, which helps with formulating appropriate queries. A {\bf version} is a checkpoint of the project; in our case, this refers to a physical {\em commit} created via \git. \provdb has \emph{explicit versions} and \emph{implicit versions}; the former are created when a user explicitly issues {\em commit} command, whereas the latter are created at provenance ingestion time when the user runs commands in the project directory.\eat{A {\em version graph} captures the parent/child relationships among the versions. }

{\bf Snapshots} are checkpointed versions of an artifact, and capture its evolution. \provdb monitors file changes during the lifecycle, and emits changed (CUD) artifacts as new snapshots, and the previous snapshot of the same artifact before the change is called its \emph{parent}. The content of a snapshot are modeled as {\bf records}, to allow fine-grained record-level provenance (some files, e.g., binary files, would be modeled as having a single record).

{\bf Derivations} capture the transformation context to the extent possible. If a derivation is performed by running a program or a script, then the information about it is captured along with any arguments. Derivation edges may also be created when the system notices that one or more artifacts have changed (e.g., an edit made using an editor, or a script ran outside the \provdb context).

Finally, {\bf properties} are used to encode any additional information about the snapshots or the derivations, as {\em key-value} pairs (where values are often time series or JSON documents themselves). Provenance ingestion tools (discussed in next section) will generate these properties. In addition to the information about programs or scripts and their arguments, properties may include any information captured by parsing shell scripts or analysis scripts themselves.  Properties are also used to extract statistics about the data within the snapshots as well, so that they can be seamlessly queried. This starts \ul{blurring the distinction between data and metadata} to some extent; we plan to investigate using a more elaborate data model that more clearly delineates between the two in future.

\topicul{Physical Property Graph Data Model:} We map the logical data model (with the exception of {\bf Record}) above into a property graph data model, primarily to enable graph traversal queries and visual exploration over the stored information easily. Nodes of the property graph are of types {\em Version}, {\em Artifact}, etc., whereas the edges capture the parent-child relationships (Figure~\ref{fig:data_model_all} shows an example).

\huicomment{a formal definition of provenance may be needed.}

\huicomment{file view needs to be reflected}

\huicomment{parameter search needs to be reflected}

\section{Provenance Ingestion}
\label{sec:ingestion}

\gdoccomment{Describes 3-4 different mechanisms to capture provenance.
* Using the DLV commandline language, which allows searching for parameters etc., and which allows running scripts with parameters.
* Provide the ability to associate a provenance capture script with a type of file, and allow the ability to construct provenance using that (proposal has some discussion).
* Manually add provenance information.
* Read stuff from git.}

\provdb captures provenance information or other metadata opportunistically, and features a suite of mechanisms that can capture provenance for different types of operations. Users can easily add provenance ingestion mechanisms, to both capture more types of information as well as richer information. Here we briefly enumerate the ingestion mechanisms that \provdb currently supports, which include a general-purpose UNIX shell-based ingestion framework, ingestion of DVCS versioning information, and a mechanism called {\bf file views} which is intended to both simplify workflow and aid in fine-grained provenance capture.

\begin{comment}
Although a typical DVCS user can capture historical snapshots of artifacts with parental relationships between versions, the detailed changes between two commits are missing. For example, the transient parameter used to produce artifacts, temporal order and dependencies between artifacts, and not-committed intermediate snapshots.

To collect fine-grained provenance as much as possible, \provdb takes over users shell commands, commits implicitly at each command (\emph{One-Command-One-Commit}), and ingests the changes \emph{before/during/after} the commands run.  In order to not pollute user commits, user explicit commits are annotated, and \provdb git index is self-contained.
\end{comment}

\topicul{Shell command-based Ingestion Framework}: 
The provenance ingestion framework is centered around the UNIX commandline shell (e.g., bash, zsh, etc). We provide a special command called {\tt provdb} that users can prefix to any other command, and that triggers provenance ingestion. Each run of the command results in creation of a new {\em implicit} version, which allows us to capture the changes at a fine granularity. These implicit versions are kept separate from the explicit versions created by a user through use of {\tt git commit}, and are not visible to the users. A collection of {\em ingestors} is invoked by matching the command against a set of regular expressions, registered a priori along with the ingestors. \provdb schedules ingestor to run before/during/after execution the user command, and expects the ingestor to return a JSON property graph consisting of a set of key-value pairs denoting properties of the snapshots or derivations. An ingestor can also provide {\em record-level provenance} information, if it is able to generate such information.

A default ingestor handles abitrary commands by parsing them  following  POSIX standard (IEEE 1003.1-2001) to annotate utility, options, option arguments and operands. For example, {\tt mkdir -p dir} is parsed as utility \emph{mkdir}, option \emph{p} and operand \emph{dir}. Concatenations of commands are decomposed and ingested separately, while a command with pipes is treated as a single command. If an external tool has been used to make any edits (e.g., a text editor), an implicit version is created next time {\tt provdb} is run, and the derivation information is recorded as missing.

\begin{comment}
In a high level, an ingestor defines a invocation filter using regular expression to match the command line. \provdb identifies the ingestors, pass the command line to the ingestor. An ingestor \emph{before/during/after} command module is called accordingly, and \provdb expects a return of JSON property graph consist of a set of key-value properties about files or derivations. If external tools instead of command line is used (e.g. text editor), any missing provenance can be identified if \emph{before} command call finds changed files comparing with previous implicit commit.

As arbitrary commands can be run, a user plugin system is provided. The user can add invocation filters to execute a shell program (plugin) in her preferable language. The plugin sees the current snapshots, as well as changed file list, and outputs the ingested provenance property graph as JSON. It is worth mentioning, the plugin can be an executable in the current project, which itself has provenance.
\end{comment}

\provdb also supports several specialized ingestion plugins and configurations to cover important data science workflows. In particular, it has an ingestor capable of ingesting provenance information from runs of the {\tt Caffe} deep learning framework; it not only ingests the learning hyperparameters from the configuration file, but also the accuracy and loss scores by iteration from the result logging file.\eat{(this is an example where {\em properties} are populated by analyzing the contents of a file).} We are currently working on incorporating parsers for scripts written in popular data science tools such as {\tt Jupyter}, {\tt scikit-learn} and {\tt pandas}, by building upon prior work~\cite{noworkflow}.

\topicul{User Annotations}: 
Apart from plugin framework, \provdb GUI allows users to organize, add, and annotate properties, along with other query facilities. The user can annotate project properties\eat{  is allowed to add provenance information manually}, such as usage descriptions for collaborations on artifacts, or notes to explain rationale for a particular derivation. A user can also annotate a property as parameter and add range/step to its domains, which turns a derivation into a template and enables batch run of an experiment. For example,  a grid search of a template derivation on a start snapshot\eat{with a start snapshot, a template derivation, a grid search for an operand,} can be configured directly in the UI. Maintaining such user annotations (and file views discussed next) as the datasets evolve is a complicated issue in itself~\cite{katz1990toward}.

\topicul{File Views}: 
\provdb provides a functionality called \emph{file views} to assist dataset transformations and to ingest provenance among data files. Analogous to views in relational databases, a file view defines a virtual file as a transformation over an existing file. A file view can be defined either: (a) as a script or a sequence of commands (e.g., {\tt sort | uniq -c}, which is equivalent to an aggregate count view), or (b) as an SQL query where the input files are treated as tables. For instance, the following query counts the rows per label that a classifier predicts wrongly comparing with ground truth.
\\{\small{
\verb|provdb fileview -c -n='results.csv' -q='|\\
\verb|  select t._c2 as label, count(*) as err_cnt |\\
\verb|  from {testfile.csv} as t, {predfile.csv} as r|\\
\verb|  where t._c0 = r._c0 and t._c2 != r._c2 group by t._c2'|\\
}}
%It can be used with command suite {\tt provdb fileview}.
\eat{
It can be used with {\tt provdb} commandline suite to create a fileview.
\\{\small{
\verb~  provdb fileview [-c|-e|-l|-d] [-n=<name>] [-q=<sql>]~\\
}}
}
The SQL feature is implemented by loading the input files into an in-memory {\tt sqlite} database and executing the query against it. Instead of creating a view, the same syntax can be used for creating a new file instead, saving a user from coding similar functionality.% in a script.

File views serves as an example of a functionality that can help make the ad hoc process of data science more structured. Aside from making it easier to track dependencies, SQL-based file views also enable capturing record-level provenance by drawing upon techniques developed over the years for provenance in databases. % (\provdb currently does not support this).

\begin{comment}
Current practice of data transformation involves in complex scripting using proprietary libraries or writing one time utility develop by the user. 
\squishlist
\item In \provdb, several continuous derivation steps can be marked as a \emph{file view}, once it is marked, a virtual file handler is allowed to be used in command arguments. A retrieval on \emph{file view} uses latest snapshots of artifacts to apply the derivations in its order. 
\item For more structured data (e.g. csv), \provdb provides SQL like transformation language instead of letting the user write scripts. Running a transformation query is also a derivation. Not only a \emph{file view} can be similarly created, but also the record level fine-grained provenance can be captured using the well defined semantics of SQL operators.
\squishend
\end{comment}

\topicul{Discussion:} Current \provdb prototype is designed to be used in a command-line environment. In future work, we plan to investigate how to provide tools for capturing provenance within other development environments such as different IDEs. We also plan to incorporate support for ingesting provenance through parsing log files generated in many environments today, and through continuous monitoring of the artifacts in the working directory.

\begin{figure*}[t!]
\centering{
\subfigure[Diff Artifacts (Result logging files for two deep neural networks)]{
  \includegraphics[width=0.48\textwidth]{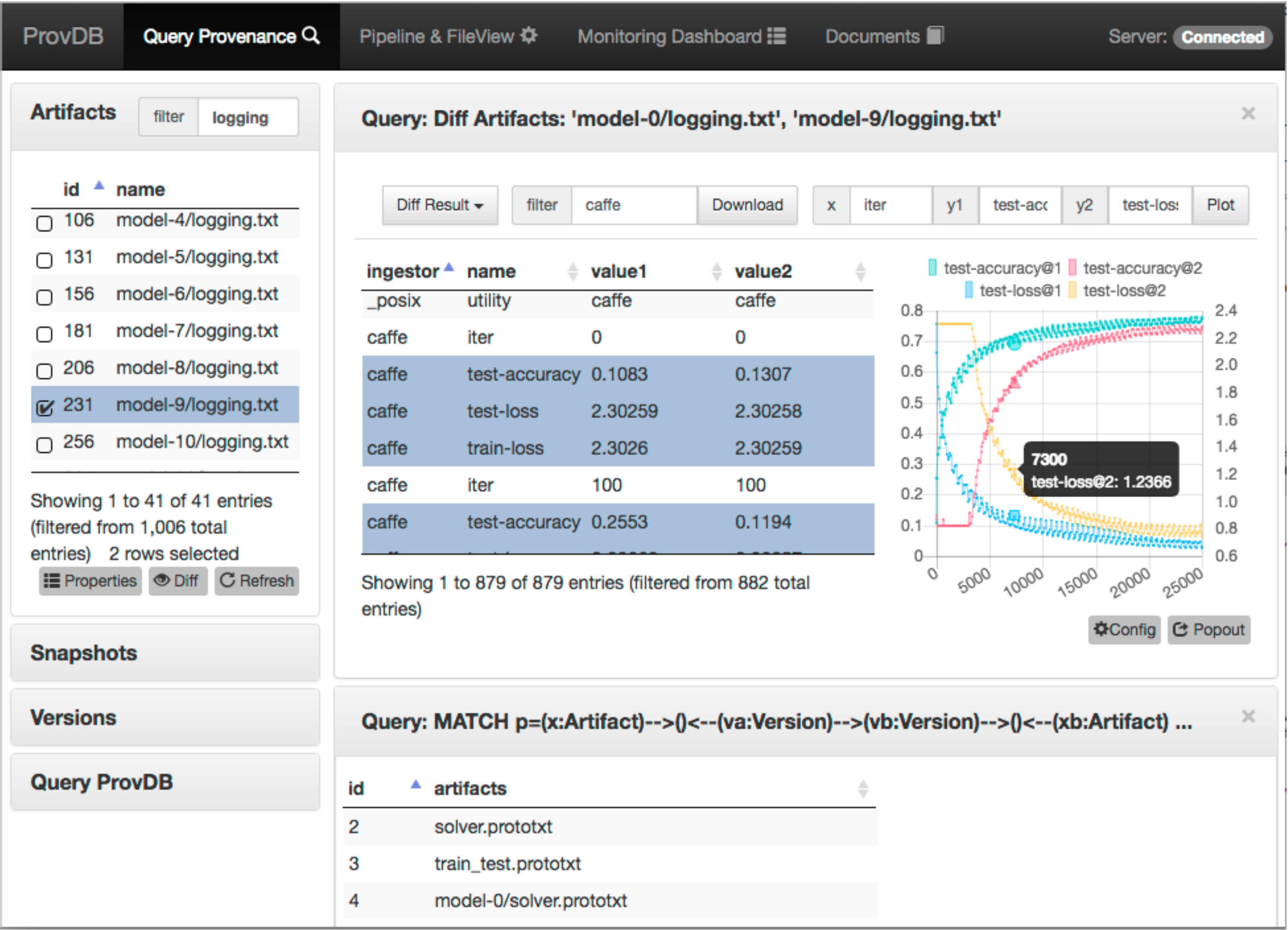}
  \label{fig:showcase_diff_nav}
}
~~
\subfigure[Cypher Query to Find Related Changes via Derivations]{
  \includegraphics[width=0.48\textwidth]{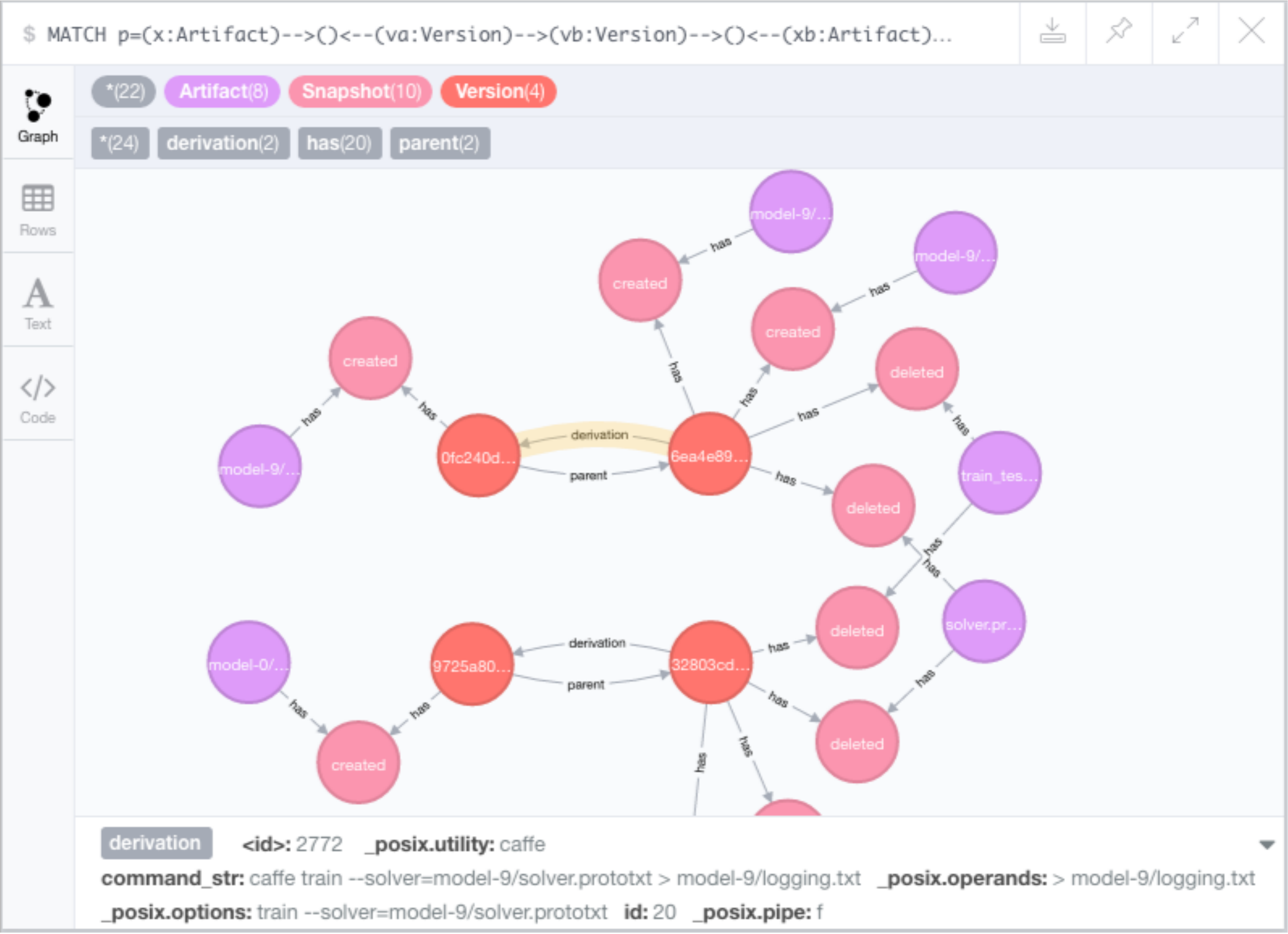}
  \label{fig:showcase_neo_nav}
}
}
%\vspace{-10pt}
\caption{Illustration of \provdb interfaces}
%\vspace{-15pt}
\label{fig:showcase}
\end{figure*}

\eat{
\begin{figure}[t!]
\subfigure[Diff Artifacts (Result logging files for two deep neural networks)]{
  \includegraphics[width=0.48\textwidth]{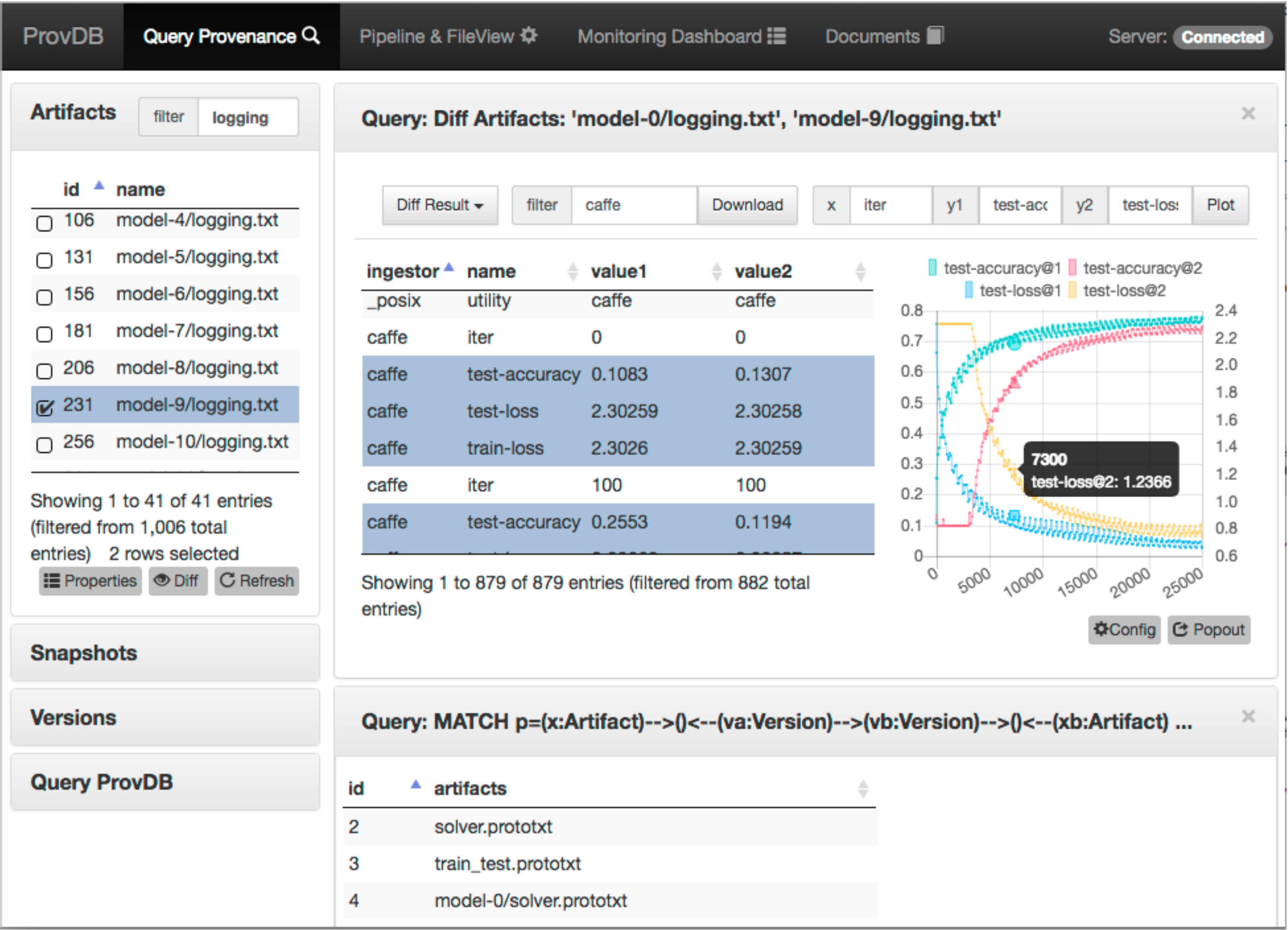}
  \label{fig:showcase_diff_nav}
}
\subfigure[Cypher Query to Find Related Changes via Derivations]{
  \includegraphics[width=0.48\textwidth]{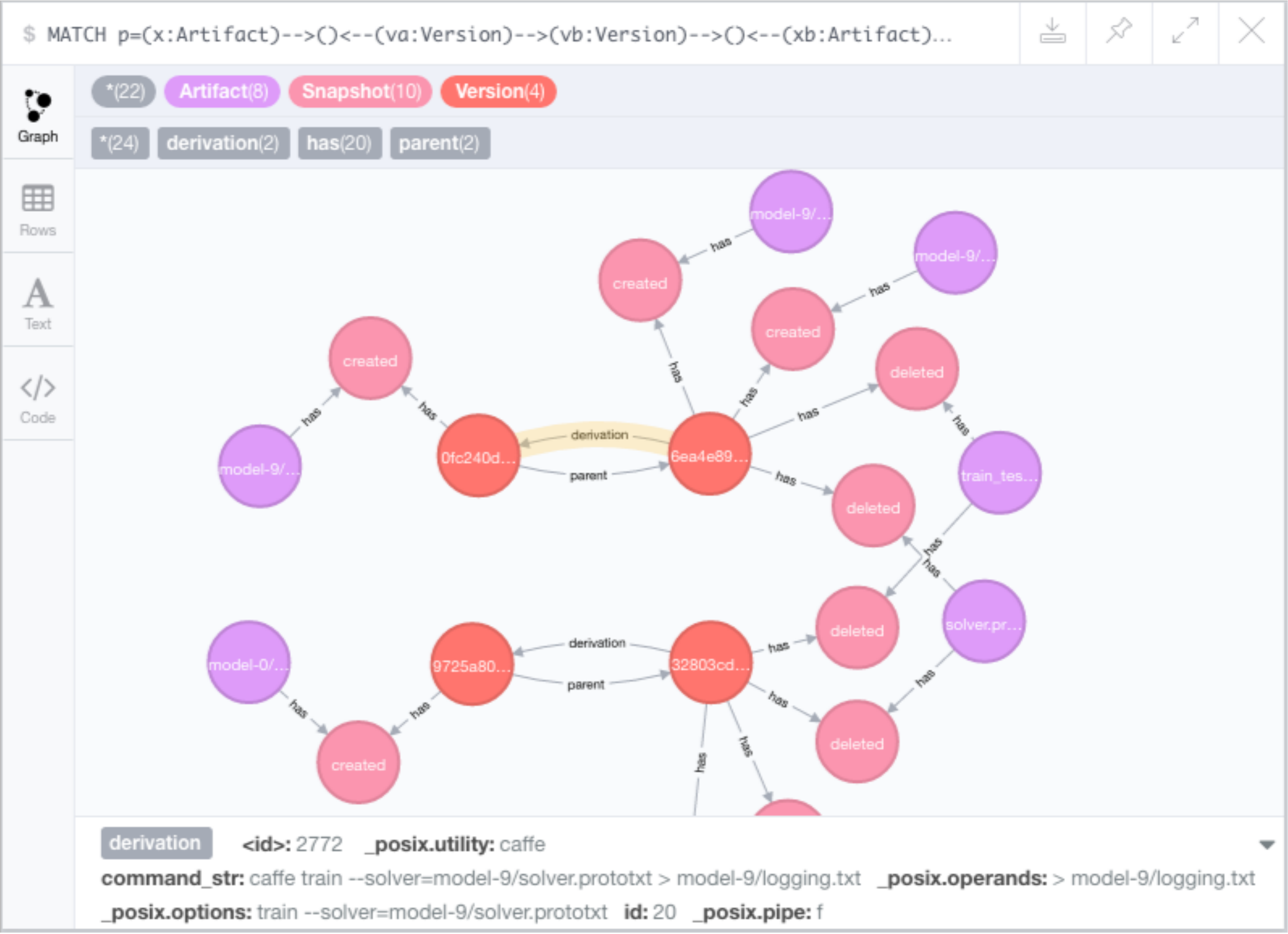}
  \label{fig:showcase_neo_nav}
}
%\vspace{-10pt}
\caption{Illustration of \provdb interfaces}
%\vspace{-12pt}
\label{fig:showcase}
\end{figure}
}

\eat{\section{Querying and Analyzing Provenance/Metadata Information}}
\section{Query and Analysis Facilities}
\label{sec:query}

\gdoccomment{Describe some of the types of interesting queries we support.
* Simple graph queries over the data (using Gremlin or CYPHER or something).
* Point to two result relations and ask what are the differences between them.
* Aggregate TaPP-style queries, evaluated in a dumb fashion (by checking out everything)}

The major data management research challenges in building a system like \provdb revolve around querying, analyzing, and extracting insights from the rich provenance information collected using the mechanisms described so far. In addition to standard provenance queries, \provdb also enables asking deeper, introspective questions about the data science processes and pipelines, and formalizing those is a major challenge in itself. \provdb can also naturally support {\em monitoring} queries, which can be used to automatically detect problems during deployment. We hope that building the basic infrastructure to collect and expose the information will allow other researchers and data scientists to start formulating such questions more easily. Developing a higher-level query language also remains a major challenge; although we proposed an initial design of a query language in our prior work~\cite{provenance2015}, it does not support querying over workflow derivations or analysis artifacts.

One major challenge that we do not discuss further here has to do with efficiency; in real deployments with large collaboration teams, a large amount of provenance information may be collected that will likely overwhelm a single, centralized instance (especially if record-level provenance information is collected). New techniques for compressing the information, and maintaining the information in a distributed manner would need to be developed.

In the rest of this section, we briefly discuss the different types of queries or analyses that may be performed on the provenance information, and what our current prototype supports.

\begin{comment}
Under \provdb property graph data model with DVCS lineages and detailed derivation context, \provdb naturally supports coarse provenance on file level entities and fine-grained provenance on records. Moreover with derivations and user ingested properties, \provdb can answer meaningful queries fitting project with different stereotypes (e.g. data cleaning, prediction, deep learning, etc.). \provdb supports \huicomment{query language: cypher? sql? vquel? or new dsl?} to query the relationship and metadata collected. Moreover, \provdb GUI allows related queries and navigations to explore the project provenance. We categorize and describe the queries as follows, and leave the GUI part in Sec.~\ref{sec:demo}:
\end{comment}

\eat{\topic{Queries over Versioning/Derivation Information and Properties}: }
\topicul{Queries over Version/Workflow Graph and Properties}: 
In a collaborative workflow, provenance queries to identify what revision and which author last modified a line in an artifact are common (e.g., {\tt git blame}). \provdb allows such queries at various levels (version, artifact, snapshot, record) and also allows querying the properties associated with the different entities (e.g., details of what parameters have been used, temporal orders of commands, etc). In fact, all the information exposed in the property graph can be directly queried using the Neo4j Cypher query language, which supports graph traversal queries and aggregation queries.

\begin{comment}
\provdb enrich the basic DVCS query in two ways:
\squishlist
\item Version graph queries at Version/Artifact/Snapshot/Record different levels can be answered. For example, find the version information (e.g. user, date) where an artifact first appeared, find the subset of artifacts often changed all together in history, and find dependencies among artifacts.
\item How the modifications are made can be queried using the derivations. As derivation is a command history and ingested properties that command run on snapshots, the user can examine the details conducted by peers, for instance, what parameters have been used, whether the temporal orders of commands are misused so that the result is wrong.
\squishend  
\end{comment}

%\topic{Derivation \& Property Queries for Different Workflows}

The latter types of queries are primarily limited by the amount of context and properties that can be automatically ingested. \provdb currently supports ingestors for several popular frameworks, including a program analysis ingestor for \emph{scikit-learn} which extracts the scikit-learn APIs used in a program, and a hyper-parameter and result-table ingestor for \emph{caffe} for deep learning (the hyper-parameter ingestor extracts experiment parameter metadata from \emph{caffe} commands and arguments, while the results-table ingestor extracts optimization errors and accuracy metrics from training logs). Availability of this information allows users to ask more meaningful queries like: what scikit-learn script files contain a specific sequence of commands; what is the learning accuracy curve of a caffe model artifact; enumerate all different parameter combinations that have been tried out for a given learning task, and so on.

Many such queries naturally result in one or more time series of values (e.g., properties of an artifact over time as it evolves, results of ``diff'' queries discussed below); \provdb supports a uniform visual interface for plotting such time series data, and comparing two different time series (see below for an example).

\topicul{Shallow vs Deep ``Diff'' Queries:}
``Diff'' is a first-class operator in \provdb, and can be used for finding differences at various different levels.  
Specifically, given a pair of nodes (corresponding to two snapshots) in the property graph, a {\em shallow} diff operation, by default, focuses on the ingested properties of the two snapshots, which are likely to contain the crucial differences in most cases. It attempts to ``join'' the two sets of properties as best as it can, and highlights the differences; in case of time-series properties, it also allows users to generate plots so they can more easily understand the differences. For example, for two {\em result table artifacts} that may represent the outputs of two different runs of the same script (e.g., model training logs), a line-by-line diff may be useless because of irrelevant and minor numerical differences; however, by plotting the two sets of results against each other, a user can more quickly spot important trends (e.g., that a specific value of parameter leads to quicker convergence). The shallow diff operator also allows differencing the contents of the two files line-by-line if so desired.

A {\em deep} diff compares the ancestors of the two target snapshots by tracing back their derivations to the common ancestor. It aligns the snapshots along the two paths, and shows the differences between each pair of aligned snapshots. For example, in a prediction workflow, a user may have tried out different prediction models and configurations to identify the best model; using \provdb, she can start from two result table artifacts, and ask a \emph{deep diff} query to compare how they are derived.

\begin{comment}
Furthermore, an user provided ingestor can always produce complex diff results as derivation properties, share with others, and query them later. 
\end{comment}

\topicul{Record Provenance Queries:} Although the \provdb data model supports storing fine-grained record-level provenance information, it currently does not have an ingestor that generates such data; we are working on adding several such ingestors, including ones for SQL-based file views or transformations, and for common data cleaning or similar operations where record-level provenance can be easily inferred. Given such information, record-level provenance queries are conceptually straightforward. However, the main challenge is expected to be the large volume of provenance information as well as efficient query execution. We plan to investigate approximate (lossy) provenance storage mechanisms to address these challenges. The utility of these queries may also be limited because it is difficult to collect fine-grained provenance for many black-box operations (e.g., machine learning models). Developing techniques to do so remains a rich area for further work.

\topicul{Reasoning about Pipelines:} Similar to a workflow management system, we define a pipeline to be a sequence of derivation edges. A pipeline can be annotated by the user by browsing the workflow graph and marking the start and the end edges of the pipeline. Pipelines can also be inferred automatically by the system (e.g., via pattern mining techniques). \provdb UI allows a user to browse and reuse pipelines present in the system; in future, we also plan to add support for re-invoking an old pipeline on an old artifact to verify the results, or invoking a pipeline on a different snapshot with different parameters, or schedule a cron job. Being able to reason about pipelines has the potential to hugely simplify the lives of data scientists, by allowing them to learn from others and also helping them avoid mistakes (e.g., omission of a crucial intermediate step).

\topicul{Continuous Monitoring or Anomaly Detection:} 
On ingested properties of artifacts and derivations, \provdb provides a monitoring and alerting subsystem to aid the user during the development lifecycle. We envision two main use cases for this functionality. (a) First, it can be used to detect any major changes to the properties of an evolving dataset -- e.g., a large change in the distribution of values in a dataset may be cause for taking remedial action. (b) Second, in most applications, there is usually a need to ``deploy'' an analysis script or a trained model against live incoming data; it is important to keep track of how well the model or the script is behaving and catch any problems as soon as possible (e.g., changing input data properties; higher error rates than expected). Currently even if systems like Spark Streaming or Apache Storm can be used to execute a script against new data in a streaming fashion, there is no built-in support for the introspection tasks. Newer systems like Google TensorFlow Serving also facilitate the deployment process, but do not support introspection. Such introspection can be seen as continuous queries against streaming provenance information.

\provdb supports analysis of historical data (as described above) and simple alert queries that can monitor a property of an evolving artifact. In our current prototype, both of these must be done through the web dashboard UI; in future iterations of \provdb, we plan to support more complex temporal queries (that can monitor properties across snapshots) and we plan to support executing those continuously as new versions (implicit or explicit) are checked in.

\topicul{Illustrative Example:} Figure~\ref{fig:showcase} shows the \provdb Web GUI using a {\tt caffe} deep learning project. In this project, 41 deep neural networks are created for a face classification task. The user tries out models by editing and training models. In Fig~\ref{fig:showcase_diff_nav}, an introspection query asks how different are two trained models (\emph{model-0} and \emph{9}). Using the GUI, the user filters artifacts, and diffs their result logging files. In the right side query result pane, the ingested properties are diffed. The {\tt caffe} ingestor properties are numerical time series; using the provided charting tool, the user plots the training loss and accuracy against the iteration number. From the results, we can see that \emph{model-9} does not train well in the beginning, but ends up with similar accuracy. To understand why, a deep diff between the two can be issued in the GUI and complex Cypher queries can be used as well. In Fig.~\ref{fig:showcase_neo_nav}, the query finds previous derivations and shared snapshots, which are training config files; more introspection can be done by finding changed hyperparameters.

\begin{comment}
\gdoccomment{We have to decide if it makes sense for this to be a demo paper rather than a research paper. We should probably have a visualization in any case, that allows someone to explore the provenance informatio and do some simple things with it.}

\provdb provides a set of interfaces for the users to ingest provenance and introspect workflow. The user can also use Neo4j interface to explore property graphs if necessary.

\topic{Command line suite}: {\tt provdb} command line suite is able to control version and manage nodes ({\tt init, commit, tag}), ingest provenance for a shell command ({\tt ingest}), issue graph query using cypher ({\tt query}), create and execute fileviews ({\tt fileview}) and interact with the html interface ({\tt gui}).

\topic{HTML GUI}: \provdb comes with a set of templated queries and navigations that a user often uses in a local HTML GUI. The main features include: a) list/search/plot properties of nodes, b) traverse neighbors of a selected node via derivation edges, c) diff/visualize node/edge properties, d) issue abitrary cypher query, e) configure and view monitoring dashboard, f) list and execute fileviews.

\topic{Showcase}:
\end{comment}

\section{Conclusion}
In this paper, we presented our vision for a system to simplify lifecycle management of ad hoc, collaborative analysis workflows that are becoming prevalent in most application domains today. We argued that a large amount of provenance and metadata information can be captured passively, and analyzing that information in novel ways can immensely simplify the day-to-day processes undertaken by data analysts. We have built an initial prototype using {\tt git} and Neo4j, which provides a variety of provenance ingestion mechanisms and the ability to query, analyze, and monitor the captured provenance information. Our initial experience with using this prototype for a deep learning workflow (for a computer vision task) shows that even with limited functionality, it can simplify the bookkeeping tasks and make it easy to compare the effects of different hyperparameters and neural network structures. However, many interesting and hard systems and conceptual challenges remain to be addressed in capturing and exploiting such information to its fullest extent.

%\small
\bibliographystyle{abbrv}
\bibliography{datahub-refs,newrefs}

%\bibliographystyle{abbrv}
%\bibliography{datahub-refs,newrefs}

%\input{data-management.tex}
%\input{personnel.tex}
%\input{collaborations-and-collaborators.tex}

\end{document}